# Tableaux accessibles dans les documents numériques

Katerine ROMEO[1], E. PISSALOUX[2], F. SERIN[3]

[1]IUT de Rouen, [2]Université de Rouen Normandie, [3]IUT du Havre

*Laboratoire LITIS/CNRS FR 3638*

**THEMATIQUE** – Informatique – Robotique – Imagerie

**RESUME** - *L'accessibilité des tableaux sur les sites web pour les personnes présentant une incapacité visuelle (PPIV) n'est pas optimale avec les lecteurs d'écran qui ne sont pas toujours performants pour la restitution d'informations visuelles (2D). Les technologies web/multimédia actuelles ne prennent pas en compte le décalage de la perception visuelle par rapport à la perception vocale qui est linéaire. Ce papier analyse les difficultés d'accès à l'information spatiale avec les recommandations existantes pour la conception des sites web accessibles à tous. Différentes solutions pour faciliter la création des tableaux accessibles sont présentées.*

*Mots-clés*— **E-accessibilité, tableaux HTML, lecteur d'écran, déficience visuelle.**

1. INTRODUCTION

Une étude sur la 'hiérarchie perçue' des handicaps dans l'entreprise, montre que les employés pensent qu'une personne non voyante est considérée comme pouvant rencontrer des difficultés très importantes à assez importantes, si elle travaillait dans une entreprise. Le pourcentage des opinions dans ce sens atteint 85% et figure au premier rang de tous les handicaps dans l'entreprise [1]. Ce pourcentage tombe à 46% quand le salarié interrogé travaille au quotidien avec des personnes en situation de handicap. Néanmoins, ce pourcentage reste au premier rang par rapport aux autres handicaps sensoriels ou mentaux. Travailler avec des personnes atteintes de cécité est donc perçue comme difficile par plus d'une personne sur deux. Le travail devenant de plus en plus numérique, nous sommes face à un problème d'accessibilité à l'information des documents numériques avec des images, des graphiques et des éléments perçus en 2D comme des tableaux.

Selon l'OMS (Organisation Mondiale de la Santé) 217 millions de personnes présentent une déficience visuelle modérée à sévère, dans le monde, dont 36 millions sont touchées par la cécité ou par une malvoyance suffisamment forte pour les astreindre à une assistance autre que l'agrandissement des informations via une loupe [2]. En France ce nombre atteint respectivement 380000 et 60000. Malgré les progrès de la médecine et de la prévention, cette prévalence reste constante par le fait du vieillissement des populations et l'allongement de la vie.

Cet article n'est pas une nouvelle communication chargée d'alerter sur les soucis d'accessibilité. Nous avons voulu proposer une approche qui permette aux contributeurs d'appréhender la complexité d'accès aux documents numériques mais aussi de proposer une solution qui facilite de bonnes pratiques. Nous nous concentrons dans cet article, sur les tableaux de données réalisés au format HTML, pour montrer que leur structure et leur mise en place peut être complexe pour leurs auteurs et surtout pour les rendre accessibles auprès d'un public cible : les personnes présentant une incapacité visuelle (PPIV). Si les normes et les lois incitent fortement à l'accessibilité, la conception d'un code adapté ne va pas de soi. Nous présenterons les outils qu'utilisent les déficients visuels pour s'approprier les contenus de tableaux et nous proposerons une solution pour que cette mise en accessibilité ne soit pas un obstacle pour les personnes qui sont amenées à créer des tableaux.

Cette communication présente la problématique d'accessibilité aux différents types de tableaux, leur usage sur le web et les faiblesses des lecteurs d'écran les plus populaires pour accéder aux informations contenues dans les tableaux (Section 2). La perception sensorielle visuelle comparée à la perception vocale est analysée du point de vue de la présentation de tableaux (Section 3). Des solutions recommandées pour une meilleure prise en charge par un lecteur d'écran sont proposées (Section 4). Le prototype développé pour faciliter la création de tableaux est présenté dans la Section 5. La Section 6 conclut et indique quelques prolongements possibles du travail présenté.

2. QU'EST-CE QU'UN TABLEAU DE DONNEES ?

Il existe une grande variété de tableaux avec des logiques et des objectifs différents. Dans notre étude, nous nous limitons à la définition standard rencontrée la plupart du temps dans des documents susceptibles d'être imprimés ou visualisés sur écran, c'est-à-dire avec une vue plane. Un tableau est alors, de façon générale, une représentation de données à deux dimensions, pour être vue d'un coup d'œil.

*2.1 Définition et usage des tableaux sur le web*

Les quatre usages des tableaux sont :

a) Ensemble de données (chiffres, chaines de caractères, dates…), des renseignements disposés de façon méthodique permettant leur comparaison ou bien constituant un regroupement de données.
b) Tableau synoptique (didactique) vue générale qui permet de saisir d'un coup d'œil les diverses parties d'un ensemble (par exemple : tableau des poids et des mesures ; liste nominative de personnes, catégories).
c) Table analytique (table de vérité, table logarithmique)
d) Tableau de mise en forme d'une page web.

Nous rappelons que l'usage des tableaux HTML pour la mise en forme d'une page Web est fortement déconseillé et dégrade l'accessibilité par les lecteurs d'écran. Cet usage qui tend à décroître reste encore présent dans les envois d'informations par courrier électronique. WCAG 2.0 (Référence d'accessibilité W3C [3], RGAA (Référentiel Général d'Accessibilité pour les Administrations) [4] et Accessiweb (Critère 5.8 et 5.3) [5] tolèrent les tableaux de mise en forme à condition de ne pas utiliser d'éléments propres aux tableaux de données et que le contenu linéarisé reste compréhensible. Le dernier point reste difficile à vérifier si le concepteur n'a pas de lecteur d'écran.

Une distinction peut également être faite entre tableaux simples et tableaux complexes. Un tableau simple comporte un seul niveau d'entêtes sur des colonnes et/ou des lignes. Il est considéré comme complexe quand il y a plusieurs niveaux et ensembles d'entêtes de lignes ou de colonnes ou s'il y a une fusion de cellules. Dans la Figure 1, on peut voir le modèle UML d'un tableau complexe. Une cellule est associée à une colonne et une ligne. En HTML, il existe des attributs **colspan** et **rowspan** qui réunissent plusieurs cellules pour former ce que nous avons qualifié dans le schéma de groupe de cellules (GroupeCell). Dans un tableau simple, tout groupe de cellule ne regroupe qu'une et une seule cellule. Un tableau peut être imbriqué dans un autre tableau mais cet usage est fortement déconseillé car il détériore grandement l'accessibilité et l'intelligibilité.

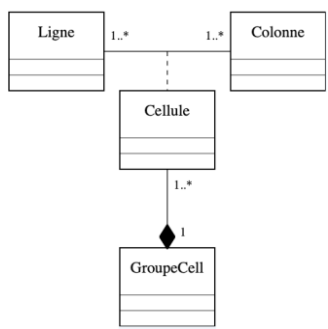

Fig.1 Modèle UML d'un tableau complexe

*2.2 Rendu des lecteurs d'écran*

Un lecteur d'écran est un logiciel qui permet de lire le contenu d'un site grâce à une synthèse vocale mais aussi de transmettre ce contenu à un afficheur braille. Il existe plusieurs lecteurs d'écran parmi lesquels les plus populaires sont Voiceover (installé sur iPhone, iPad et les ordinateurs Apple par défaut), NVDA (Non Visual Desktop Access, développé en 2007 en open source) et JAWS (Job Access With Speech, produit commercial développé en 1989). Le principe de ces lecteurs d'écran est d'accéder au DOM (Document Object Model) et d'utiliser les interfaces de programmation du navigateur pour avoir les informations sur les objets, leur nom, leur état et leur rôle dans le but de les annoncer au moment du parcours de la page. Grâce aux balises sémantiques HTML (correctement utilisées), la synthèse vocale indique l'organisation spatiale.

3. PERCEPTION SENSORIELLE : VISUELLE VS VOCALE

L'exploration des documents numériques sollicite naturellement la perception visuelle car ces documents sont conçus par des personnes sans incapacité visuelle. L'organisation des documents vise à être attractive pour attirer le regard et disperser l'information sur le support numérique. Les lecteurs d'écran linéarisent le contenu parcouru (text-to-speech) et utilisent le balayage séquentiel. Cette lecture linéaire est évidemment une contrainte pour les tableaux où le regard amène la structure planaire. Nous pouvons ajouter aussi que pour une perception cognitive de l'information en 2D, il est nécessaire de faire des retours en arrière horizontalement et verticalement de façon la plus fonctionnelle possible. Cette stratégie de parcours d'un tableau conditionne la charge cognitive et la vitesse d'accès à l'information.

4. RENDRE ACCESSIBLE UN TABLEAU DE DONNEES

*4.1. Tableaux en HTML*

Les tableaux sont apparus dès 1994 grâce au navigateur Netscape et les balises furent intégrées au standard HTML en 1997. Les quatre balises fondamentales sont : <table> indique le début et la fin du tableau, <tr> ligne, <th> entête de ligne ou de colonne, <td> donnée. Les balises qui sont apparus par la suite : <caption> qui indique le titre du tableau, <thead> qui regroupe les entêtes du tableau, <tbody> l'ensemble de données, <tfoot> qui indique souvent la ligne des totaux par colonne ou des informations liées aux données du tableau.

Tableau 1. Exemple d'un tableau complexe.

| Pays | Grandes Villes | | Habitants (millions) |
|---|---|---|---|
| | Capitale | Métropole | |
| **Algérie** | Alger | | 34 |
| **Australie** | Canberra | Sydney | 42 |
| **Belgique** | Bruxelles | | 12 |
| **Brésil** | Brasilia | Sao Paulo | 110 |

*4.2. Critères d'accessibilité*

Dans un document non-linéaire, tel un tableau, l'attention du lecteur se retrouve face à un contexte élargi de l'exploration de l'information. Sa représentation mentale devient plus délicate par le passage à une description linéaire 1D pour une phrase à une structure 2D plus ou moins complexe. La PPIV est privée de la perception visuo-spatiale. Le lecteur d'écran apporte souvent une première information d'importance dès qu'un tableau est abordé : la taille de celui-ci. Le tableau représenté sera labellisé par une légende ainsi que sa dimension (nombre de colonnes et de lignes) selon le critère d'accessibilité 223 sur le site Opquast [6] et Accessiweb [5] (critères 5.4 et 5.5). La balise **caption** devra être renseignée. Sur la balise **table**, il est possible aussi d'ajouter un attribut **title**, pour indiquer des informations supplémentaires sur le contenu des colonnes et la logique d'organisation du tableau. Accessiweb conseille l'utilisation de l'attribut **summary** pour un résumé concis du tableau, ce qui amène souvent la répétition des informations par les concepteurs. Par ailleurs summary est considéré comme obsolète par HTML5 dans l'utilisation des tableaux [7]. La balise **details** peut être utilisée imbriquée dans une balise caption. Une autre solution serait d'insérer l'élément table dans un élément **figure** en ajoutant la description textuelle dans un élément **figcaption**.

Les cellules des tableaux de données sont reliées à leurs entêtes (Critère 222, Opquast), (Critère 5.7, Accessiweb), (H63, H43, WCAG 2.0) ce qui est réalisé par l'attribut **scope** dans les balises **th**, en indiquant les en-têtes des lignes (row) ou des colonnes (col) que le lecteur lit pour chaque cellule pour rappeler l'information spatiale des cellules dans le tableau. Scope peut prendre **colgroup** ou **rowgroup** comme valeur si les entêtes sont fusionnés. Dans ce cas l'attribut **colspan** ou **rowspan** est ajouté pour indiquer combien de cellules sont fusionnées verticalement ou horizontalement. La propriété scope est considérée comme obsolète en HTML5 mais elle reste quasi indispensable pour les lecteurs d'écran actuels. C'est pour le moment la seule technique efficace et partagée pour restituer certaines informations [7]. L'attribut **abbr** est utilisé quand le titre de la colonne ou ligne est long et qu'une abréviation s'avère utile pour la lecture de chaque cellule.

Une autre alternative pour que le lecteur d'écran rappelle le contexte spatial est l'utilisation des attributs **id** et **headers**. Dans ce cas **id** est ajouté à tous les en-têtes en spécifiant par un mot leur contenu. Ces identifiants de lignes et de colonnes sont repris à la lecture des cellules. Si le tableau est complexe, les cellules fusionnées peuvent être lues avec les entêtes indiqués dans l'attribut **headers** de ces cellules.

WAI-ARIA [8] conseille d'ajouter l'attribut **role** dans la balise **tr** et si c'est nécessaire l'attribut **aria-describedby** dans les attributs **th**. Des répétitions sont observées avec ce dernier attribut, ce qui suggère de ne l'utiliser qu'en cas d'indications précises qui n'ont pas été dites par ailleurs [9].

*4.3. Exemple d'accessibilité des tableaux sur Wikipédia*

L'accessibilité des tableaux est, en général, assez peu traitée dans les faits. Nous prendrons comme exemple le site de Wikipédia. C'est la plus grande encyclopédie. Elle est alimentée quotidiennement par plus de cent mille contributions à travers le monde. Elle est visitée chaque mois par près de 500 millions de visiteurs (Juin 2017). Wikipédia est le site le plus visité en France qui ne soit ni un moteur de recherche, ni un réseau social, ni un site marchand, avec 24 millions de visiteurs uniques par mois. Sur Wikipédia, il existe trois documents qui expliquent la façon de créer des tableaux : débutants (~250 mots), avancé (~1600 mots), expert (~6500 mots). Ce n'est qu'au niveau expert que les problèmes d'accessibilité sont abordés. L'atelier dédié à l'accessibilité ne compte pas moins de 10000 mots ; la partie dédiée aux tableaux en compte environ 1800. On comprendra donc qu'à moins de pratiquer volontairement l'accessibilité, celle-ci est loin d'être aisée. Le codage pratiqué par MédiaWiki (moteur de Wikipédia) est, de plus, très proche de l'encodage HTML ce qui ne le rend compréhensible que de développeurs avertis [10].

Il existe une autre forme de tableaux, les plus répandus sur Wikipédia : les tableaux pré-dimensionnés pour recevoir des informations présentes dans les bases de données de l'encyclopédie. Il s'agit du format Infobox. Dans les documents expliquant le fonctionnement de ces tableaux, le terme accessibilité n'est employé que pour aborder l'accessibilité aux données. C'est-à-dire le contrôle de ces données ; il n'est pas fait allusion à une politique d'accessibilité orientée vers le handicap. Les exemples donnés contiennent des titres cachés non accessibles par les PPIV.

5. PROTOTYPE PROPOSE

Il est possible de simplifier la conception de tableaux via un CMS (Content Management System) tout en lui donnant une accessibilité étendue. Nous avons développé un logiciel en Java qui prend à la base un texte simple et linéaire comme le font les éditeurs de type Wikitext. Il n'exige pas de connaître et encore moins d'introduire les balises et les propriétés (attributs) imposées pour améliorer l'accessibilité telles que celles proposées par le WAI-ARIA. Notre approche différencie trois phases dans la conception d'un tableau :

*5.1 Phase 1 : écriture d'un code simple et accessible par l'usager rédacteur.* Les données sont rentrées avec un séparateur laissé au choix du contributeur : pipe (ligne verticale), point-virgule ou tabulation. La fusion de cellules est possible sur une ligne (colspan). Le choix est possible de multiplier à volonté le nombre de lignes en en-têtes voire de ne pas en indiquer. Afin de permettre la séparation entre lignes d'en-tête et corps d'un tableau, il est nécessaire d'indiquer (par une suite de tirets) la séparation physique des deux parties. Une mise en place automatique des références lignes/colonnes est faite.

*5.2 Phase 2 : analyse du code pour permettre la conception d'un tableau.* Notre analyseur va déterminer la dimension et la structure de l'en-tête. Cette partie est modélisée sous forme de segments, ils seront autant de références utiles pour renseigner les cellules lors du parcours par le lecteur d'écran. La décomposition en colonnes du corps du tableau va permettre, ensuite, de lier chaque cellule à un ensemble d'informations présentes dans l'entête. En cas d'usage de la propriété **colspan**, le logiciel sait fusionner les indications pour rappeler la fusion (Fig.1). Les références utiles à chaque colonne sont détectées par une progression descendante, permettant de placer les informations de la plus générale à la plus spécifique.

*5.3 Phase 3 : réalisation en HTML du tableau par notre générateur en garantissant l'inclusion des balises ARIA ad hoc.* Les parties **thead** et **tbody** sont placées dans les tableaux. Le premier recueille les lignes d'en-tête tandis que le second contient les lignes avec les données. Ces deux parties permettent de signifier la sémantique qui différencie en-tête et valeurs. Chaque cellule fera donc explicitement référence aux cellules d'en-tête avec la propriété headers suivie des identifiants générés lors de la création des en-têtes. La propriété **scope** précise pour chaque cellule son appartenance à la ligne identifiée par la cellule indiquée par la balise **th**.

6. CONCLUSIONS ET PERSPECTIVES

Cette communication a proposé une analyse de l'accessibilité des tableaux sur le web et dans les documents numériques en général. Nous avons mis en évidence quelques critères importants dans la présentation et utilisation des tableaux. Nous avons développé un prototype pour faciliter l'accessibilité des tableaux dans un CMS. Nous sommes en train de faire une série de tests auprès d'un échantillon d'usagers PPIV ou non.

Un standard de structuration logique du contenu des tableaux devrait être guidé par sa représentation interne balisée, et tenir compte des mécanismes cognitifs de recherche des informations [11][12].